# Engineering ultrafast spin currents and terahertz transients by magnetic heterostructures

T. Kampfrath[1], M. Battiato[2], P. Maldonado[2], G. Eilers[3], J. Nötzold[1], I. Radu[4], F. Freimuth[5], Y. Mokrousov[5], S. Blügel[5], M. Wolf[1], P. M. Oppeneer[2], M. Münzenberg[3]

1. Department of Physical Chemistry, Fritz Haber Institute, Berlin, Germany.
2. Department of Physics and Astronomy, Uppsala University, Uppsala, Sweden.
3. I. Physikalisches Institut, Georg-August-Universität Göttingen, Göttingen, Germany.
4. Helmholtz-Zentrum Berlin für Materialien und Energie, Berlin, Germany.
5. Peter Grünberg Institute and Institute for Advanced Simulation, Forschungszentrum Jülich and JARA, Jülich, Germany.

**In spin-based electronics, information is encoded by the spin state of electron bunches[1,2,3,4]. Processing this information requires the controlled transport of spin angular momentum through a solid[5,6], preferably at frequencies reaching the so far unexplored terahertz (THz) regime[7,8,9]. Here, we demonstrate, by experiment and theory, that the temporal shape of femtosecond spin-current bursts can be manipulated by using specifically designed magnetic heterostructures. A laser pulse is employed to drive spins[10,11,12] from a ferromagnetic Fe thin film into a nonmagnetic cap layer that has either low (Ru) or high (Au) electron mobility. The resulting transient spin current is detected by means of an ultrafast, contactless amperemeter[13] based on the inverse spin Hall effect[14,15] that converts the spin flow into a THz electromagnetic pulse. We find that the Ru cap layer yields a considerably longer spin-current pulse because electrons are injected in Ru d states that have a much smaller mobility than Au sp states[16]. Thus, spin current pulses and the resulting THz transients can be shaped by tailoring magnetic heterostructures, which opens the door for engineering high-speed spintronic devices as well as broadband THz emitters[7,8,9], in particular covering the elusive range from 5 to 10THz.**

Contemporary electronics is based on the electron charge as information carrier whose presence or absence encodes the value of a bit. Much more efficient devices for low-power data storage and processing could be realized if the spin degree of freedom were used in addition[1,2,3,4]. The spintronics approach requires the generation and control of spin currents, that is, the transport of spin angular momentum through space[5,6]. Spintronic operations should be performed at a pace exceeding that of today's computers, which ultimately requires the generation of spin current pulses with terahertz (1 THz = $10^{12}$ Hz) bandwidths[7,8] as well as the possibility to manipulate them in novel structures[17,18]. To date, femtosecond spin-current pulses have been successfully launched by optically exciting electrons in semiconductors[10] or ferromagnetic metals[11,12]. However, to enable ultrafast basic operations on these transients (such as buffering or delaying), their shape and propagation have to be controlled on sub-picosecond time scales.

Here, we employ magnetic heterostructures containing an optimally chosen nonmagnetic metallic layer whose electron mobility allows us either to trap or to transmit electrons and, thus, to engineer ultrafast spin pulses. The spin flow is probed in a contactless manner using the inverse spin Hall effect[14,15] (ISHE) that converts the spin current into a detectable THz electromagnetic pulse[13]. Our findings open up a route to device-oriented femtosecond spintronics as well as novel broadband emitters of THz radiation[7,8,9].

Our idea is illustrated in Fig. 1a, which shows a schematic of a ferromagnetic Fe film capped by a thin layer of Ru or Au. Absorption of a femtosecond laser pulse (photon energy 1.55eV) in the Fe layer promotes electrons from below the Fermi energy to bands above it (Fig. 1b), thereby generating a non-equilibrium electron distribution. The photoexcited majority-spin electrons in Fe (spin up in Fig. 1) have mainly sp-like character and a ~5 times higher velocity than the d-type excited minority-spin electrons[19] (spin down in Fig. 1). As a consequence, transport of spin polarization from the Fe into the cap layer, that is, a spin current, will set in immediately[11,12,20]. We expect a very different transport dynamics in the Fe/Au and Fe/Ru structures because Au has a much higher electron mobility than Ru (Ref. 16). On a microscopic level, the non-equilibrium electrons arriving in the Au layer will occupy only sp states that exhibit high band velocity[19] (~1nm fs$^{-1}$) and a long lifetime[19] (~100fs). The electrons will thus reside in the Au layer for a relatively short while before they are reflected back into the Fe film. In Ru, on the other hand, the hot electrons will mainly occupy more localized d-band states with a lower band velocity[21] (~0.1nm fs$^{-1}$), and they undergo more scattering, in part because of the stronger electron-phonon coupling[16]. Therefore, transport of the non-equilibrium electrons should occur much slower in the Ru than in the Au layer, accompanied by significantly more spin accumulation.

To quantify the suggested spin-trapping scheme, we have modeled the situation of Fig. 1a with superdiffusive spin transport (see Ref. 20 and Methods). This model allows us to predict the change $\Delta \boldsymbol{M}(z,t)$ in the sample magnetization (spin density) and the associated spin current density $\boldsymbol{j}_s(z,t)$ as a function of the position $z$ perpendicular to the film plane (Fig. 1a) and the delay time $t$ since laser excitation. The pump pulse (duration 20fs, photon energy 1.55eV, absorbed energy 1.3mJ cm$^{-2}$) has constant intensity throughout the depth of the 12nm thin heterostructure (Fig. 1a). As seen in Fig. 2a, there is equilibrium ($\Delta \boldsymbol{M}=0$) at negative times. Upon optical excitation at $t=0$, the situation changes drastically, and we observe a nearly instantaneous spin transport from the Fe film towards the cap layer. In the case of Ru, the spin angular momentum is transferred until $t \approx 300$fs, after which a transient equilibrium is established. Note that we obtain strikingly different dynamics for the Au cap layer: after a slight spin accumulation in the Au, the magnetization flows back into the Fe, and within ≈300fs, the system has nearly returned to equilibrium. The very different spin accumulation in the Ru and Au cap layers is consistent with the expectation that the Ru d electrons propagate much slower than the Au sp electrons, leading to an enhanced spin trapping and slower dynamics in Ru. The system will return to $\Delta \boldsymbol{M}=0$ on much longer time scales of several 10ps by spin flips and subsequent magnon generation[22], which is not modeled here.

To put our theoretical results to test, we perform an experiment according to the scheme shown in Fig. 1. The sample has the geometrical parameters as depicted in Fig. 1a and consists of polycrystalline Fe, Ru or Au, evaporated onto a glass substrate. The magnetization of the sample is in-plane and single-domain, and its direction is set by a permanent external magnetic field (magnitude 80mT). A pump pulse (parameters as in the calculations) is directed onto the sample where it launches a spin current pulse with current density $\boldsymbol{j}_s(z,t)$ (Refs. 11, 12, 20). In order to probe $\boldsymbol{j}_s$, we borrow concepts from measurement schemes of static spin currents that take advantage of the ISHE[14,15]. As indicated in Fig. 1a, the moving electrons are subject to spin-orbit coupling that deflects spin-up and spin-down electrons in different directions[23] (Fig. 1a). The spin current $\boldsymbol{j}_s$ is thus transformed into a perpendicular charge current

$$\boldsymbol{j}_c = \gamma \boldsymbol{j}_s \times \boldsymbol{M}/|\boldsymbol{M}|, \qquad (1)$$

where the spin-Hall angle $\gamma$ is a measure of the electron deflection[15,24] (Fig. 1c). Whereas in static experiments the resulting charge current is measured as a voltage[14,15,24], we here electrooptically sample the field of the electromagnetic pulse that is emitted by the charge current burst $\boldsymbol{j}_c$ (see Fig. 1a and Methods). From equation (1) and Fig. 2, we expect linearly polarized radiation with frequencies covering the THz window.

In our experiment, the emitted THz pulses are always found to be polarized parallel to the *x* direction (Fig. 1a), that is, perpendicular to the sample magnetization ***M*** pointing along *y*. Figure 3a shows measured THz signals *S*(*t*) which equal the *x* component of the transient THz field ***E***(*t*) directly after the sample (Fig. 1a), convoluted with the response function of our setup[13]. Each transient is fully inverted when ***M*** is reversed, which demonstrates a strong connection between THz emission and magnetic order. Note that we observe strikingly different THz signals from the Ru- and Au-covered Fe films (Fig. 3a). First, by extracting the THz field $E_x(t)$ from the raw data *S*(*t*) (see Supplementary Information) and by Fourier transformation, we find that the amplitude spectrum $|E_x(\omega)|$ from the Fe/Ru sample covers the frequency range from $\omega/2\pi=0.3$ to about 4THz (Fig. 3b). In sharp contrast, the emission spectrum from the Fe/Au sample extends to frequencies as high as 20 THz. Such different bandwidth indicates a much faster dynamics in the Fe/Au than in the Fe/Ru bilayer. Second, the THz waveforms have opposite sign for times *t*<1ps (Fig. 3a). Finally, the THz emission from both samples exhibits a completely different dependence on the pump-pulse energy (Fig. 3b, inset). While the energy $\int dt\ S(t)^2$ of the emitted THz pulse saturates at an absorbed excitation fluence of 0.8 mJ cm$^{-2}$ for the Fe/Ru bilayer, it still increases nearly linearly for Fe/Au.

The emitted THz transient ***E***(*t*) is a result of the *z*-averaged charge current density $\langle \boldsymbol{j}_c \rangle = \int_0^d dz\, \boldsymbol{j}_c/d$ flowing in the plane of the *d*=12nm thin heterostructure. Both quantities are connected by Ohm's law (see Supplementary Information)

$$\langle \boldsymbol{j}_c(\omega)\rangle = -\langle \sigma(\omega)\rangle \boldsymbol{E}(\omega), \qquad (2)$$

where $\langle \sigma(\omega)\rangle$ is the *z*-averaged THz conductivity. By applying equation (2) to ***E***($\omega$) (Fig. 2b), we obtain the experimental charge current transients $\langle j_{c,x}(t)\rangle$ shown in Fig. 4a. For the Fe/Au bilayer, $\langle j_{c,x}(t)\rangle$ changes sign already 300fs after its onset, which indicates a backflow of charge. Note, however, the much slower current dynamics in the Fe/Ru structure in which a charge backflow starts only when charge transport in the Au-capped sample is half finished. These observations match our expectation that spins are temporarily trapped in the Ru cap layer, for a duration of about 1ps (Fig. 4a). A more direct comparison of experiment and theory is obtained by considering the *z*-averaged spin current $\langle j_{s,z}(t)\rangle$ as derived from our superdiffusion simulation (Fig. 2). We emphasize that theory ($\langle j_{s,z}\rangle$, Fig. 4b) qualitatively reproduces all features found in the experiment ($\langle j_{c,x}\rangle$, Fig. 4a). Such agreement is consistent with our interpretation that the emitted THz transient is a result of a laser-driven spin-current pulse and its transformation into a transverse charge current. As discussed below, this point of view is supported by several consistency checks.

First, the emitted THz pulse is found to be independent of the pump polarization direction (Supplementary Fig. S1a). This observation is in line with the assumption that the spin current is carried by laser-heated electrons. By rotating the sample magnetization ***M*** within the film plane, we observe a concurrent rotation of the polarization plane of the linearly polarized THz transient (Supplementary Fig. S1b). The directional dependence agrees with that expected from the ISHE [equation (1)]. We estimate the mean spin Hall angle $\gamma = \langle j_{c,x}\rangle/\langle j_{s,z}\rangle$ [see equation (1) and Fig. 1c] using the peak values of the measured $\langle j_{c,x}\rangle$ (Fig. 4a) and the simulated $\langle j_{s,z}\rangle$ (Fig. 4b). We infer $\gamma \sim +10^{-3}$ for the Fe/Au and $\gamma \sim -10^{-3}$ for the Fe/Ru heterostructure. These values can be compared to direct *ab initio* calculations of the spin Hall angle (see Methods and Supplementary Information). For hot electrons in Fe, approximately 0.5eV above the Fermi energy, we obtain $\gamma \sim -10^{-3}$. As the electron deflection in Ru is negligible[25], this value is in good agreement with our results of Fig. 4. Polycrystalline Au, on the other hand, is known to have a positive $\gamma$ of the order of $10^{-2}$ and even higher[26], which overcompensates the electron deflection in the Fe. This effect explains the opposite polarity of the THz pulses emitted from the Fe/Au and Fe/Ru bilayers (Fig. 2a).

Second, we note that our pump pulse also induces an ultrafast quenching of the sample magnetic moment[27] by about 7%. This modulation leads to the emission of magnetic dipole radiation[28,29,30,31] and the generation of an additional spin current along $z$ via spin pumping[6,24]. In our experiment, both contributions are negligible. The magnetic dipole radiation has an amplitude one order of magnitude smaller than the electric dipole radiation originating from the spin transport. The spin pumping delivers a spin current that is two orders of magnitude smaller than that resulting from superdiffusion (see Supplementary Information).

Finally, our superdiffusion model suggests a significant spin accumulation in the Ru cap layer, reaching about $0.2\mu_B$ per Ru atom at an absorbed pump fluence of $1.3$mJ cm$^{-2}$ (Fig. 2a). Note that Ru d-band states up to 0.5eV above the Fermi energy can host at most $0.2\mu_B$ per spin direction and atom[32]. Therefore, further increase of the pump fluence and density of non-equilibrium electrons should result in a saturation of the spin current. In contrast, the predicted spin accumulation is reduced by one order of magnitude in the Au layer (Fig. 2b), which overcompensates the four times smaller Au electron density of states[19], such that the spin current should not saturate. These expectations are in good agreement with our experimental data (Fig. 3b, inset) where a saturation of the THz signal at an absorbed fluence of ~$1.3$mJ cm$^{-2}$ is found for the Fe/Ru structure, while this effect is much less pronounced for the Fe/Au bilayer.

In conclusion, we have demonstrated the generation and frequency tailoring of ultrafast spin currents in specifically designed magnetic heterostructures consisting of ferromagnetic Fe films and nonmagnetic cap layers. By using metallic cap layers with different carrier mobility, one can largely tune the spatiotemporal shape of the spin currents. The THz emission observed in the course of this process shows that the underlying ISHE is also operative at unprecedentedly high frequencies of up to 20THz. Our results open up a route to potential spintronic devices that manipulate the temporal shape, the delay, and the transverse deflection of spin-current bursts on femtosecond time scales. In terms of photonic applications, the Fe/Au heterostructure represents a novel and easy-to-fabricate broadband THz emitter that fully covers the record bandwidth from 0.3 to 20THz (Fig. 3b). This wide frequency window cannot be realized with crystal-based emitters as they typically suffer from emission gaps in the range from 5 to 10THz which originate from phonon resonances[7,8]. The polarization of the emitted radiation can easily be controlled by the direction of the sample magnetization.

# Methods

**Experimental details.** The Fe/Au and Fe/Ru heterostructures (Fig. 1a) are grown by electron-beam evaporation on a glass substrate under ultrahigh-vacuum conditions (base pressure $5\times10^{-10}$ mbar). Both samples show a nearly rectangular hysteresis curve with a coercive field of about 10mT. In the experiment, the sample is kept in an external magnetic field of 80mT and excited by linearly polarized laser pulses (duration 20fs, center wavelength 800nm, energy 0.3mJ, repetition rate 1kHz, beam diameter at sample 3.3mm full width of half maximum of the intensity) under normal incidence from the substrate side (Fig. 1a). The THz electric field is projected onto the *x* axis by means of a broadband wire-grid polarizer and subsequently detected by electrooptic sampling[7,8,13] using the 10-fs pulses from the seed laser oscillator. All measurements are performed at room temperature in a $N_2$ atmosphere.

**Theoretical model and calculations of spin transport.** Transport simulations are performed using the theory of superdiffusive spin transport[20]. The underlying semiclassical model treats the randomly occurring scattering events exactly, which includes the generation of electron cascades due to electron-electron scattering. The reflection at interfaces is also taken into account with a reflection coefficient that enforces the conservation of linear momentum of the electron crossing the interface.

The required energy- and spin-dependent lifetimes and velocities of nonequilibrium electrons in Fe and Au are taken from *ab initio* calculations[19]. Since *ab initio* values are lacking for Ru, this metal is modeled with the two spin channels behaving like that of the minority spin channel of Fe. This approach is justified because Ru has empty d states above the Fermi energy resembling the behavior of the minority channel of Fe. The difference between the trial and real transport properties in Ru is minor compared to the fundamental difference in the transport properties between Au and Ru, which we have checked for a range of transport properties (from Au-like to Ru-like). The trend presented in Fig. 2 is clear and stable, making our conclusions robust and independent of detailed aspects of Ru transport properties.

**Theoretical model and calculations of spin Hall angles.** The spin Hall conductivity (SHC) is calculated as a function of electron energy by means of the Kubo formula within density-functional theory using the full-potential linearized augmented plane-wave (FLAPW) program FLEUR (see http://www.flapw.de). We employ the generalized gradient approximation of the exchange correlation potential, a plane-wave cutoff at a wavevector of 85nm$^{-1}$ and the experimental lattice constant of 2.866Å. Further details on the computation are given in Ref. 25 and the Supplementary Information.

## Acknowledgements

We thank R. K. Campen, F. Giesen, and A. Melnikov for stimulating discussions. MM, GE, FF, YM, and SB acknowledge financial support by the German Science Foundation through SFB 602 and SPP 1538 (SpinCaT). FF and YM acknowledge funding by the HGF-YIG Program VH-NG-513 and computer time at the Jülich Supercomputing Centre. MB, PM, and PMO acknowledge support through the Swedish Research Council, the EU programs "Fantomas" and "FemtoSpin", and the Swedish National Infrastructure for Computing (SNIC).

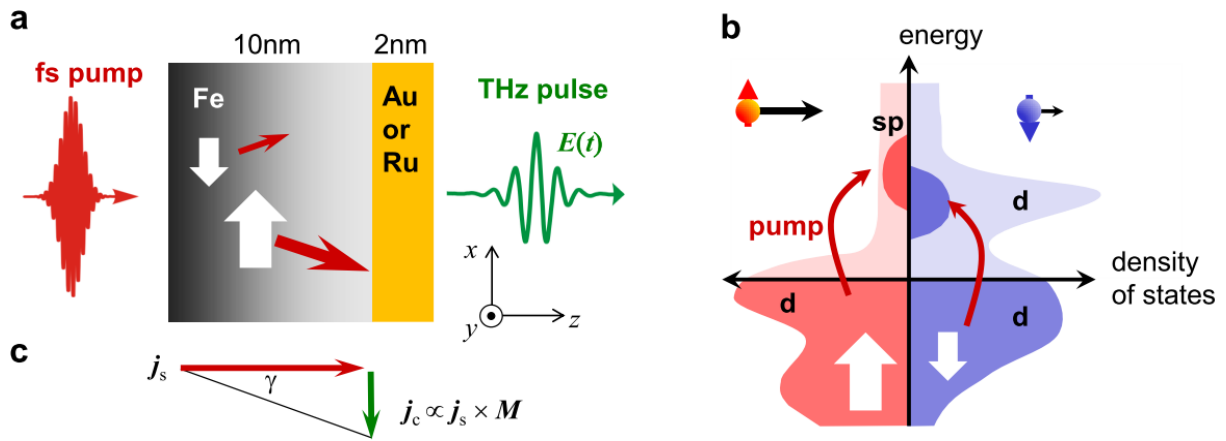

**Figure 1 | Scheme for engineering and detecting ultrashort spin-current bursts. a** A ferromagnetic Fe film (magnetization parallel to $y$ axis, perpendicular to paper plane) is excited by an optical femtosecond (fs) pump pulse, which **b** transforms slow majority-spin d electrons (red) into fast sp electrons, thereby launching a spin current towards the Au or Ru cap layer. **c** Inverse spin Hall effect (ISHE): spin-orbit interaction deflects majority and minority electrons in different directions (see **a**) and thus transforms the longitudinal spin current $j_s$ into a transverse charge current $j_c$, giving rise to the emission of a terahertz (THz) electromagnetic transient.

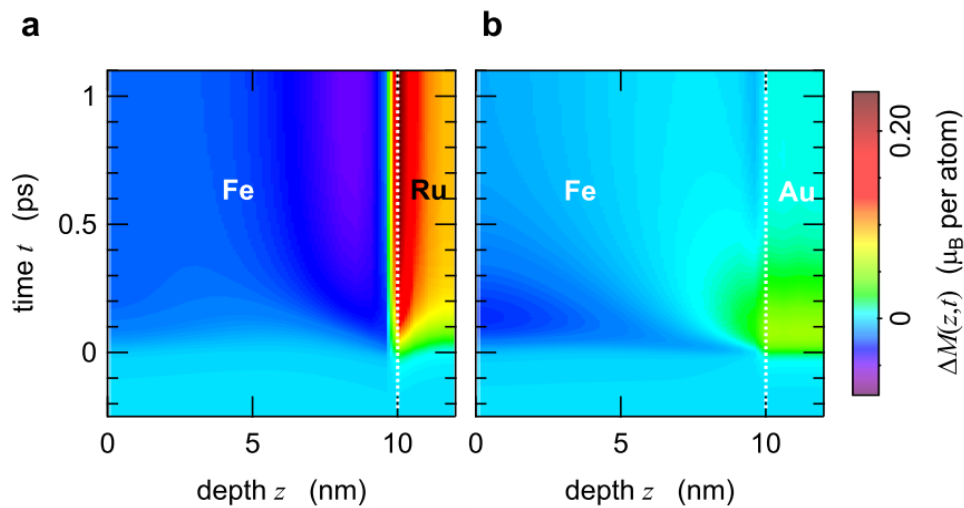

**Figure 2 | Predicted spin trapping in a magnetic heterostructure. a** Calculated magnetization change $\Delta M(z,t)$ of a Fe/Ru bilayer induced by a laser pulse (duration 20fs, absorbed pulse energy 1.3mJ cm$^{-2}$). Here, $z$ denotes in-depth film position and $t$ time since sample excitation at $t=0$. The laser intensity inside the film is nearly independent of $z$. **b** Same as **a**, yet for an Au-capped Fe film. Note the much faster dynamics and reduced spin trapping in the Au cap layer.

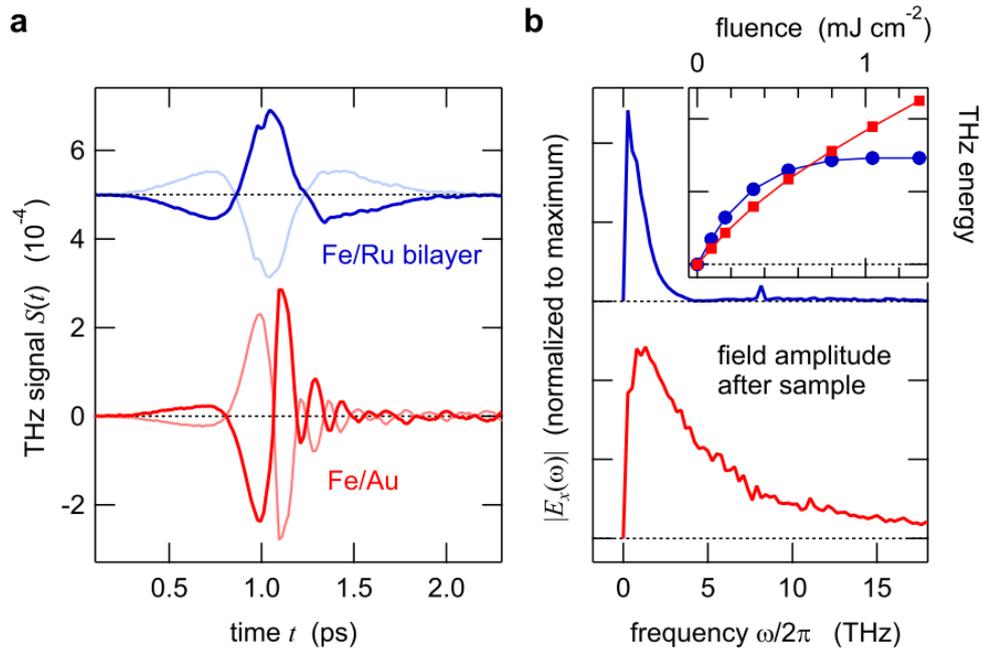

**Figure 3 | Terahertz emission from magnetic heterostructures. a** Terahertz signal waveforms obtained from photoexcited Ru- and Au-capped Fe thin films. The signal inverts with reversal of the sample magnetization (dark to light curves). Pump-pulse parameters are as in Fig. 2. **b** Fourier spectra $|E_x(\omega)|$ of the transient THz electric field directly after the sample as extracted from the raw data of **a**. Inset: emitted THz pulse energy versus absorbed pump-pulse energy per area.

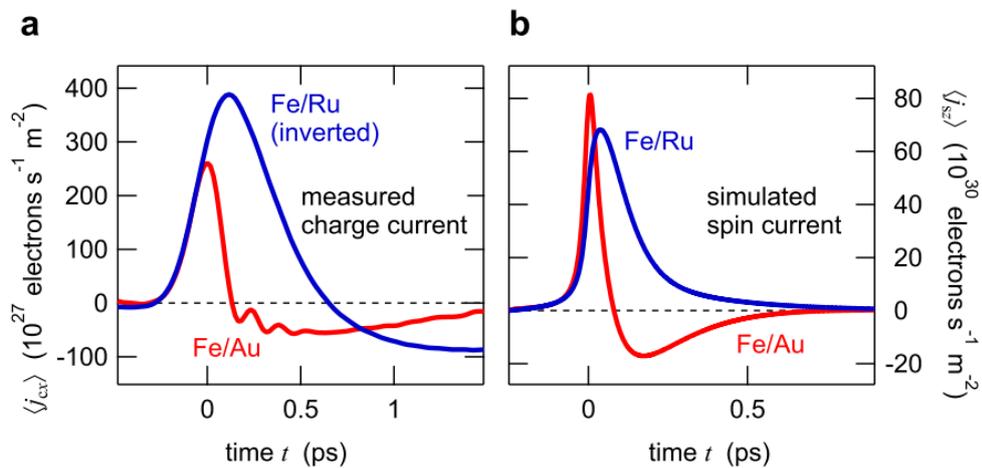

**Figure 4 | Terahertz charge and spin currents**. **a** Experimentally determined charge-current density $\langle j_{c,x} \rangle$ ($z$-averaged, flowing in-plane). The sign of the sheet current in the Fe/Ru sample has been reversed for comparison. **b** Calculated spin-current density $\langle j_{s,z} \rangle$ ($z$-averaged, flowing perpendicular to the film plane).